\documentclass[a4paper,fleqn]{cas-dc}

\usepackage[numbers]{natbib}

\usepackage{siunitx}

\usepackage{subfiles}
\usepackage{subfig}

\usepackage{adjustbox}
\usepackage{tikz}
\usetikzlibrary{positioning, arrows.meta}

\graphicspath{{figs/}}

\def\tsc#1{\csdef{#1}{\textsc{\lowercase{#1}}\xspace}}
\tsc{WGM}
\tsc{QE}
\tsc{EP}
\tsc{PMS}
\tsc{BEC}
\tsc{DE}

\begin{document}
\let\WriteBookmarks\relax
\def\floatpagepagefraction{1}
\def\textpagefraction{.001}
\shorttitle{Blender FPP Digital Twin}
\shortauthors{Weston et~al.}

\title[mode=title]{Using a Digital Twin for Fringe Projection Profilometry Optimisation}

\author{D. Weston}[
  type=editor,
  auid=000,bioid=1,
  prefix=,
  role=Researcher,
  orcid=0009-0001-4476-6556]
\ead{psydw2@nottingham.ac.uk}
\fnmark[1]

\author{X. Kong}[
  style=chinese,
  role=Researcher,
  prefix=Dr.
]

\author{G. S. D. Gordon}[
  type=editor,
  auid=000,bioid=1,
  prefix=Dr,
  role=Researcher,
  orcid=0000-0002-7333-5106]

\author{S. Piano}[
  type=editor,
  auid=000,bioid=1,
  prefix=Prof,
  role=Researcher,
  orcid=0000-0003-4862-9652]

\affiliation{organization={University of Nottingham},
  addressline={University Park}, 
  postcode={NG7 2QL}, 
  postcodesep={},
  city={Nottingham},
  country={United Kingdom}
}

\begin{abstract}
  Fringe projection profilometry (FPP) is a widely used technique for measuring object surface form and three-dimensional (3D) geometry, capable of delivering high-precision, high-resolution measurements when paired with suitable cameras and projectors. However, in practical deployments, identifying parameter configurations that maximise precision while satisfying real-world constraints remains challenging. To address this, we present an automated digital twin framework implemented in Blender, an open-source 3D software package that provides a ray-traced rendering environment that enables accurate simulation of physical systems. 
  
  We first replicated the physical setup in our digital twin by matching characterisation quality, gamma response, and characterisation images. Accurate system characterisation using Zhang's method \cite{zhang_camera_2021}, to obtain intrinsic and extrinsic parameters, is shown to be critical for achieving high precision. Using this digital twin, we then demonstrate systematic exploration and optimisation of key parameters, including phase-shift count, camera-projector spacing, and fringe density. These parameters span both system geometry (e.g. camera-projector positioning) and algorithmic choices, such as 2D phase-shifting and unwrapping methods \cite{zuo_temporal_2016}.
  
  Three measurement artefacts, representative of real-world metrology scenarios, were used to benchmark the system. The symmetrical mean Chamfer distance (SMCD), computed between ground-truth and reconstructed meshes, was used to evaluate reconstruction quality. After optimisation within the digital twin, transferring the optimal parameters to the physical system reduced the number of required images per measurement by 48\% (from 36 to 21). A reduction of 74.0\% mean SMCD was also achieved for fringe pattern stripe count alteration. A 36.9\% mean SMCD was obtained for adjusting the camera and projector spacing purely in the digital-twin. These results demonstrate that optimising an FPP system using a digital twin can significantly improve reconstruction quality while reducing measurement effort in real-world metrology applications.
\end{abstract}

\begin{graphicalabstract}
  \begin{figure*}[pos=h]
    \centerline{\includegraphics[width=\textwidth]{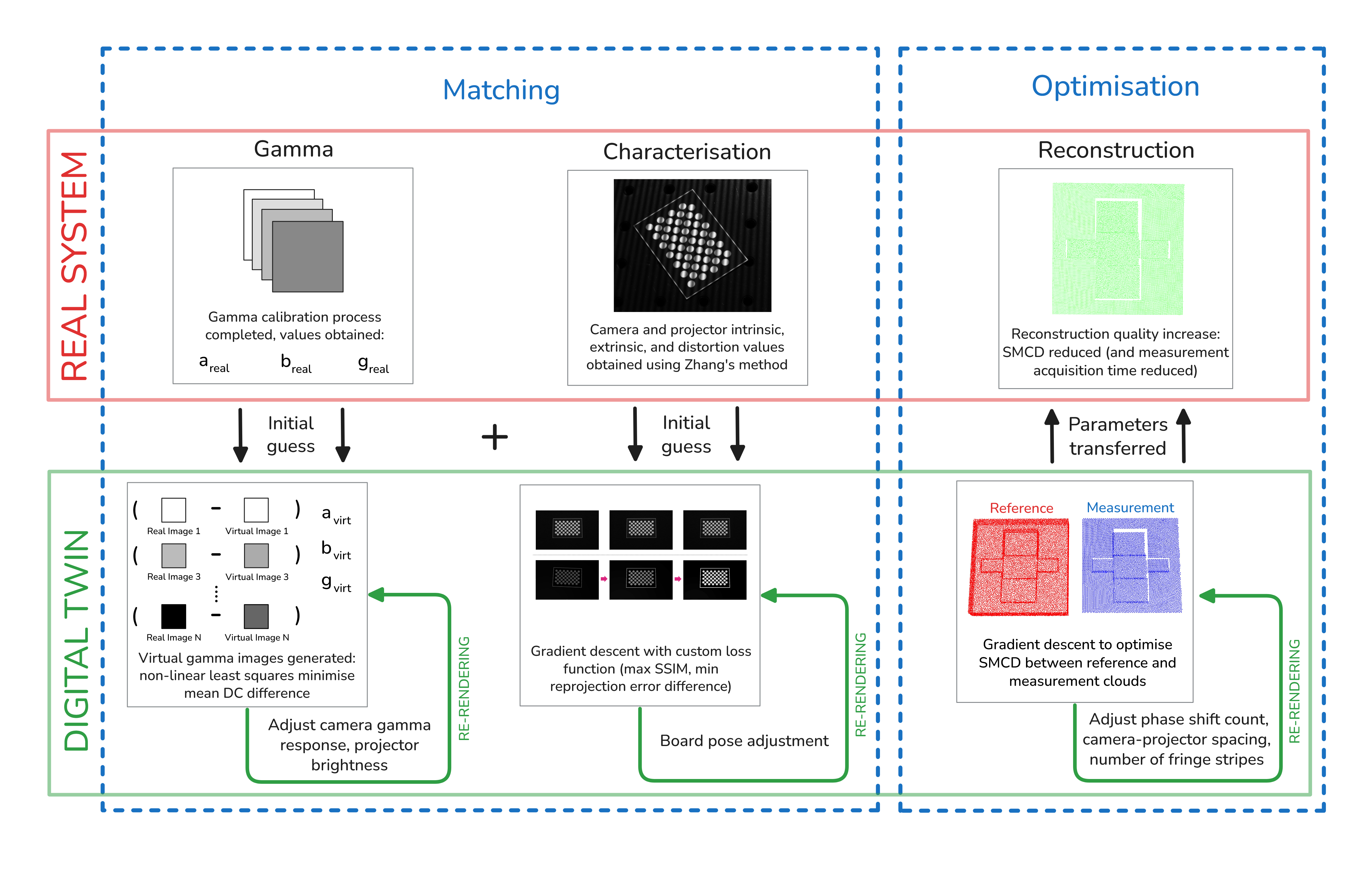}}
    \label{fig:digital-twin-overview}
  \end{figure*}
\end{graphicalabstract}


\maketitle

\section{Introduction}

\begin{figure*}[pos=h]
    \centerline{\includegraphics[width=\textwidth]{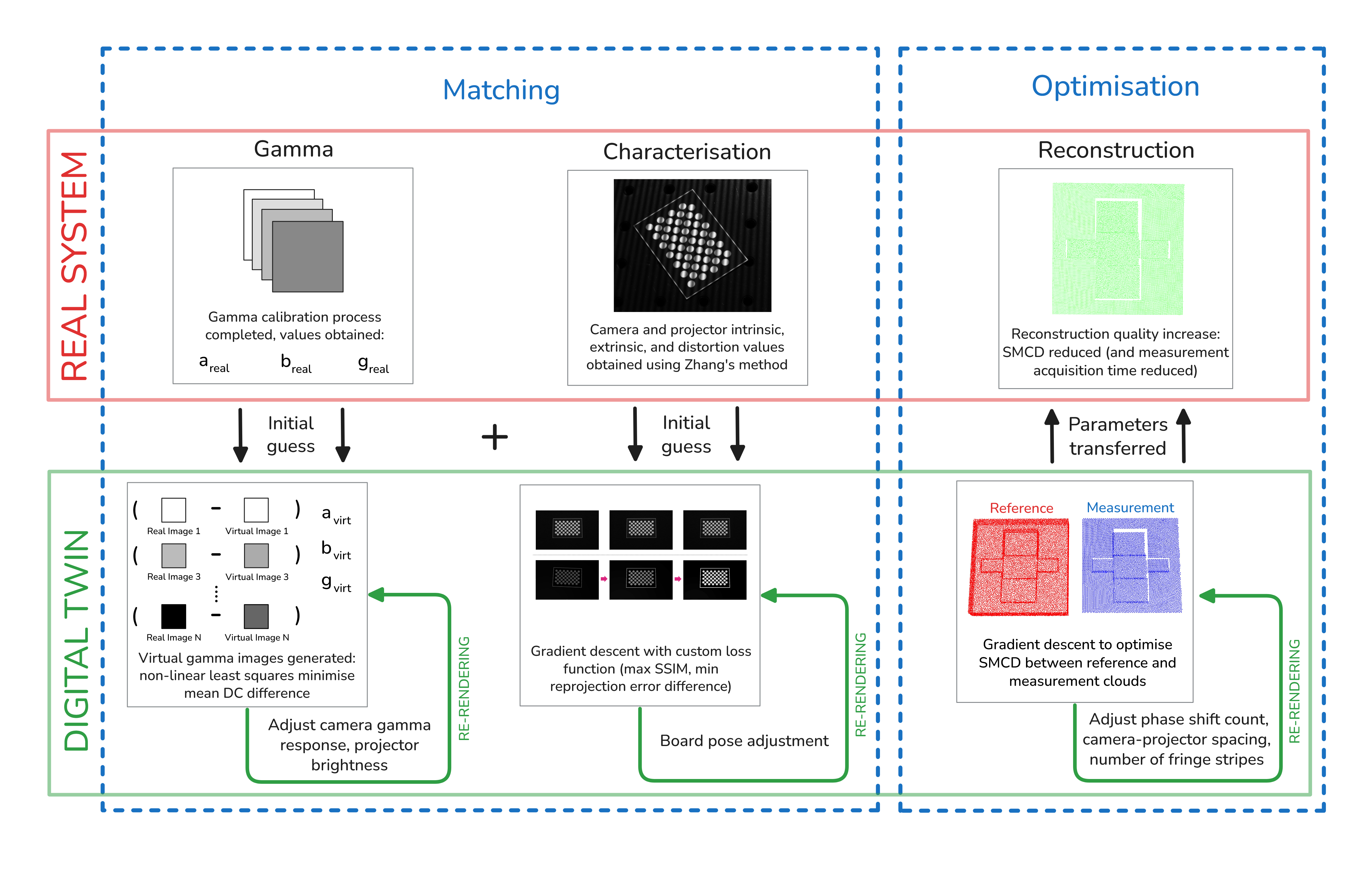}}
    \caption{Overview of the proposed real-to-digital twin framework for fringe projection profilometry. The real system involves gamma calibration, system characterisation using Zhang's method, and initial reconstruction. Transferred parameters (reference and measurement data) feed into the digital twin, where virtual gamma images are generated and optimised via non-linear least squares to minimise mean DC difference. Camera gamma response and projector brightness are adjusted using gradient descent with a custom loss function (maximising SSIM and minimising reprojection error). Board pose adjustment further optimises the cloud-to-cloud distance (SMCD) between reference and measurement clouds. Finally, system design parameters (phase shift count, camera-projector spacing, number of fringe stripes) are iteratively refined.}
    \label{fig:digital-twin-overview2}
\end{figure*}

Fringe projection profilometry (FPP) is an optical metrology technique used to measure the 3D surface form of an object using a camera and projector. Structured light patterns, typically sinusoidal fringes, are projected onto an object's surface. Surface height variations cause geometric distortions of the projected fringes, which appear as phase shifts in the observed fringe pattern by the camera. The camera records intensity images of these deformed fringes, from which the phase at each pixel is extracted and unwrapped. Using a characterised geometric model of the camera and projector, the extracted phase, at each pixel, is converted into a 3D coordinate, corresponding to a single entry in a point cloud \cite{feng_calibration_2021}.

FPP has been applied across a wide range of domains, including additive manufacturing, where it enables in-situ metrology in laser-powder bed fusion processes \cite{remani_-situ_2024}, and biomedical applications such as breast cancer detection \cite{norhaimi_digital_2020}. Owing to its flexibility and scalability, FPP has attracted significant interest across multiple disciplines, with both academic researchers and commercial manufacturers developing systems in different sizes, configurations and deployment formats \cite{hinz_fringe_2021}.

Determining an optimal trade-off between computational complexity and measurement resolution in practical systems is often laborious. Exhaustively testing large combinations of FPP configurations using physical hardware (optical devices and measurement artefacts) is time-consuming and may incur substantial financial costs. For instance, acquiring and integrating multiple cameras and projectors to empirically identify an optimal (min-max) configuration for a given application is typically impractical. Similarly, repeatedly repositioning the camera and projector is not always feasible; in many manufacturing scenarios, mechanical and design constraints fix the relative locations of these devices, preventing their adjustment after system fabrication.

Alternatively, a detailed and accurate simulation model can be used to perform this optimisation virtually. When such a model is accurately matched to the physical system with a closed feedback loop, we call this a ``digital twin''.  A related concept, a digital shadow is a virtual representation of a physical system that remains synchronised with the physical system's state without the digital twin influencing the physical system, in essence continuously ingests sensor measurements and operational data from its physical counterpart to simulate and predict the system's behaviour in real time \cite{papachristou_digital_2024}. Both digital twin and shadow approaches enable engineers and researchers to experiment with system configurations and control strategies \emph{in silico}, yielding insights that can improve the physical system's performance without risking downtime or damage. Over the past few years, digital twin technology has rapidly moved from theory to practice in domains ranging from manufacturing and aerospace to energy and healthcare \cite{attaran_digital_2023}. In advanced manufacturing, for example, digital twins are used to mirror the operation of machine tools and production lines. A notable case is the use of a CNC machine's digital twin to simulate tool paths and machining processes: by comparing the twin's simulated sensor signals with real machine data, engineers can optimise cutting parameters and detect anomalies before they lead to failures \cite{nguyen_ai-driven_2026}. 

The advantages of digital twins in such applications include the ability to perform what-if analyses, optimise system performance continuously, and react to issues proactively rather than reactively. However, there are still significant challenges to the broad implementation of digital twins. Creating and validating a high-fidelity model of a complex physical system can be extremely resource-intensive, and the accuracy of a twin hinges on the quality of real-time data and the faithfulness of the underlying models. Maintaining a tight bi-directional link between the physical and virtual systems (so that the twin stays up-to-date and can influence the physical system through controls) requires robust infrastructure \cite{nguyen_ai-driven_2026}.

There are a range of options for creating digital simulations of FPP systems, but no single, widely used software environment exists. Examples of FPP simulation tools include \cite{ribbens_projection_2013}, which was custom built, and \cite{ueda_fringe_2021}, which was developed in Unity, a game-oriented ray-tracing engine. Although \cite{ueda_fringe_2021} can produce acceptable results, their focus on using game-based rendering means it uses more limited approximations of physical environments, offers little flexibility for customisation, and does not straightforwardly render the types of variations and scenes required for metrology, which differ significantly from those in games.

These considerations led us to use Blender \cite{blender_online_community_blender_2018}, an open-source, ray-traced 3D rendering environment, to produce a simulated FPP environment. While Blender is primarily intended for ultra-realistic graphic design and animation, it provides several features that are particularly desirable for FPP simulation:

\begin{itemize}
    \setlength\itemsep{0.75em}
    \item 3D meshes (combinations of vertices, edges, and faces), and their transformations.

    \item precise light-matter interaction approximation using materials, ray-tracing (Cycles), and other techniques.

    \item performant image rendering (capture); enough data can be generated in a reasonable time-frame.
\end{itemize}

These attractive features mean that Blender has seen significant use in computer-vision tasks, including in simulating FPP systems \cite{puljcan_simulation_2022}. \cite{lemos_real-time_2025} developed a custom FPP simulator in Blender to test a novel optical-flow-based reconstruction algorithm under controlled, repeatable conditions. In the remote sensing domain, Blender has even been used to model hyper-spectral light-matter interactions; the HyperBlend project achieved leaf reflectance simulations with errors on the order of only 0.1\% compared to physical measurements \cite{riihiaho_hyperblend_2022}. These examples highlight the versatility of Blender for creating digital twins of diverse physical systems. Blender's Python scripting capabilities enable the complete automation of an end-to-end FPP system, including reconstruction algorithms. 

A virtual projector can be configured to display structured-light patterns with varying phase and spatial frequency and can optionally include noise. Cameras are available through Blender's native functionality and support configurable lens-distortion models and other properties that closely replicate real-world imaging phenomena. Objects with different textures and light interactions can be created, with configurable properties such as Lambertian reflection, light scattering, anisotropy, and normal maps that provide volumetric surface texture. These properties can be used to approximate physical objects with a high degree of accuracy, although some knowledge of the content-creation pipeline is required to obtain realistic results.

In this paper, we use Blender to create a digital twin FPP system, including a Python add-on that enables fitting a digital twin to experimental data, as well as offering end-to-end FPP experiments with algorithmic reconstruction. We first optimise the simulated system parameters to match those of a physical system by adjusting them until the simulated images closely resemble physical experimental images. We then show that this system can be used to identify optimal parameters which, when transferred back to the physical system, yield improved performance, thereby directly demonstrating the utility of this digital-twin model.

\section{Methods}
    Our work extends prior simulation-based approaches \cite{puljcan_simulation_2022} by introducing an optimisation pipeline that iteratively aligns the virtual system with the physical system using image and geometry based objective functions.

    The full process used for matching the digital twin to the physical system, referred to as ``refinement'', is shown in \autoref{fig:digital_twin_chart}.

\tikzstyle{block} = [rectangle, draw, rounded corners, text width=8cm, align=center, minimum height=1cm]
\tikzstyle{line}  = [-{Latex[length=2mm]}, thick]

\begin{figure}[pos=h]
    \centering
    \begin{adjustbox}{max width=\linewidth}
        \setlength{\fboxsep}{5pt}
        \setlength{\fboxrule}{0.5pt}
        \fbox{%
            \begin{tikzpicture}[
                block/.style={
                    rectangle,
                    draw,
                    rounded corners,
                    text width=8cm,
                    align=center,
                    minimum height=1cm,
                    inner sep=6pt
                },
                line/.style={
                    draw,
                    -{Latex[length=2mm]},
                    thick
                }
            ]
                \node[block] (n1) at (0,0) {Physical system characterised and gamma calibrated.};
                \node[block, below=0.5cm of n1] (n2) {Physical system environment (lighting, etc.) is modelled in the Blender simulation.};
                \node[block, below=0.5cm of n2] (n3) {Physical camera and projector intrinsic and extrinsic parameters are entered in the simulation.};
                \node[block, below=0.5cm of n3] (n4) {Gamma matching};
                \node[block, below=0.5cm of n4] (n5) {Characterisation matching};
                \node[block, below=0.5cm of n5] (n6) {Refinement finished};

                \draw[line] (n1) -- (n2);
                \draw[line] (n2) -- (n3);
                \draw[line] (n3) -- (n4);
                \draw[line] (n4) -- (n5);
                \draw[line] (n5) -- (n6);
            \end{tikzpicture}%
        }
    \end{adjustbox}
    \caption{Abstract view of the digital twin matching routine used to bring the two systems into close agreement.}
    \label{fig:digital_twin_chart}
\end{figure}
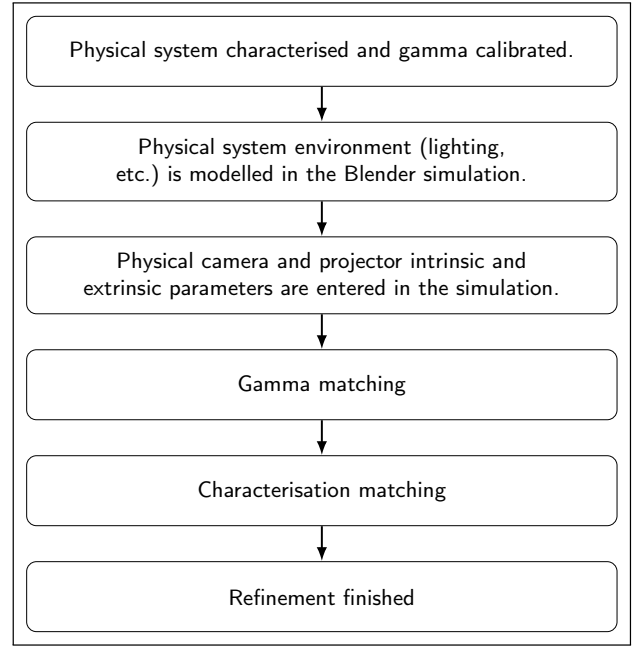

    \subsection{Fringe Projection Profilometry}
        There are many variants of FPP, for example: phase-to-height methods, Fourier transform profilometry etc. In this paper, triangular stereo FPP \cite{zhang_recent_nodate} is adopted, in which the camera and projector form a stereo pair and 3D coordinates are recovered through triangulation. This approach is widely used in existing literature, provides high-quality reconstruction, allows flexible camera-projector placement, and does not require explicit prior knowledge of their relative pose. A typical pipeline for this FPP approach is as follows:
    
        \begin{enumerate}
            \setlength\itemsep{0.75em}
            \item characterisation -- Zhang's method \cite{zhang_camera_2021} is used to estimate and refine the intrinsic and extrinsic parameters of the camera and projector,
        
            \item image acquisition and pre-processing -- fringe patterns are projected and the corresponding images are captured,
        
            \item phase extraction \cite{zuo_phase_2018} -- methods include $N$-step phase shifting, bi-frequency approaches, Fourier transform methods, and AI-based techniques,
        
            \item phase unwrapping \cite{zuo_temporal_2016,su_reliability-guided_2004} -- methods may be spatial, temporal, or AI-based,
        
            \item reconstruction \cite{feng_calibration_2021} -- a 3D point cloud is generated from the phase maps together with the intrinsic and extrinsic parameters of the system.
        \end{enumerate}
        
        The first step, characterisation, is essential in triangular stereo FPP (and more broadly metrological characterisation in general) as the intrinsic and extrinsic parameters of the measurement system must be determined in advance. The intrinsic parameters describe device-specific properties such as focal length in pixel units, lens distortion coefficients, and optical centre. The extrinsic parameters describe the geometric relationship between each measuring device and the world coordinate system, often defined relative to another device or to the characterisation board.
        
\begin{figure}[pos=h]
        	\centering
        	\includegraphics[width=.9\columnwidth]{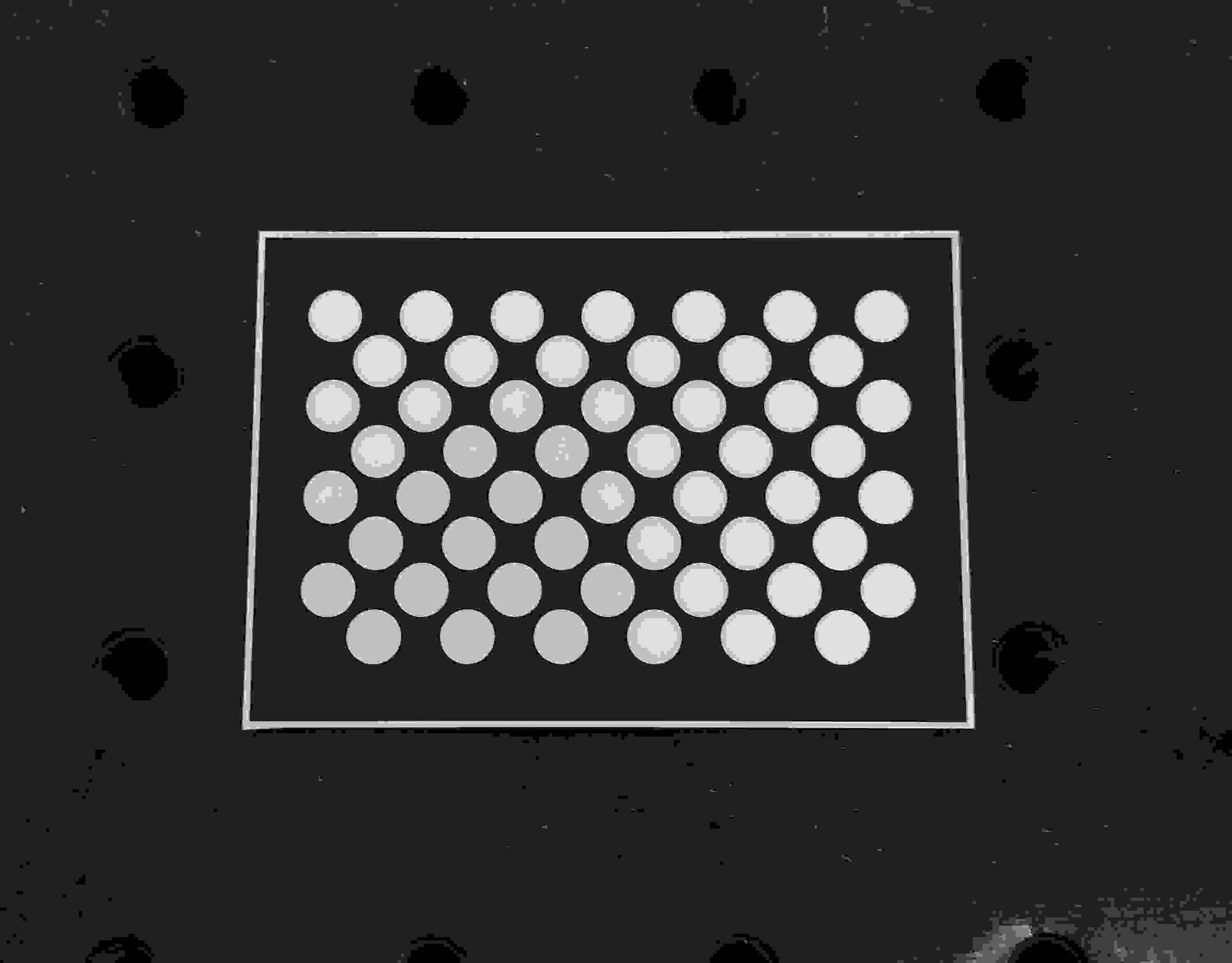}
        	\caption{The physical characterisation board used: \SI{60}{\milli\metre} width $\times$ \SI{42}{\milli\metre} height, with 52 total POIs.}
        	\label{fig:real_board}
        \end{figure}

        In this work, both camera and projector are modelled as pinhole imaging devices \cite{sturm_pinhole_2021}. In practice, optical distortion and defocus are present due to lens effects, but these can be compensated through an accurate characterisation. A common approach is Zhang's method \cite{zhang_camera_2021}, which uses a characterisation target (\autoref{fig:real_board}) containing points of interest (POIs) with precisely known coordinates. The pixel coordinates of the POIs in the captured images can then be detected using standard computer vision algorithms, depending on the POI type. For example, for circle-based POIs, the Hough Circle Transform \cite{yuen_comparative_1990} can be used to accurately identify the centre of each circle.
        
        FPP fringe patterns are projected onto the characterisation board, and images are captured for muliple poses. The phase is then extracted for each set of fringes, followed by generation of an unwrapped phase map for each image set. In this work, phase extraction and unwrapping are performed using $N$-step phase shifting and multi-frequency temporal unwrapping \cite{zuo_temporal_2016, su_reliability-guided_2004}. The projector POI coordinates can then be obtained from these images using phase matching \cite{feng_calibration_2021}.
        
        The known 3D coordinates of the POIs, together with their corresponding pixel coordinates in the camera and projector images, are used to estimate the intrinsic and extrinsic parameters via a non-linear least-squares optimisation. To ensure robustness, a diverse set of characterisation board poses (typically ten or more) is required \cite{icasio-hernandez_vdivde_2024}. In addition to geometric characterisation, other factors influencing FPP performance must be addressed, including gamma correction and fringe-direction optimisation \cite{zhang_robust_2017}. Following independent characterisation of the camera and projector, their results can be combined through joint characterisation, further improving overall system accuracy \cite{zhang_camera_2021}.

    

    \subsection{Blender Implementation}
        Blender provides access to cameras with explicit control over camera intrinsic parameters, including focal length, exposure, lens distortion coefficients, and gamma responses. 
        
        As projectors are not natively supported in Blender, a custom add-on was developed to emulate projector behaviour using Blender's shader pipeline -- the Cycles rendering engine must be used to ensure correct projector behaviour and usage of ray-tracing for rendering.

        Blender does not provide predefined mappings to physical quantities. For example, the intensity of a projector is measured in candela (or lumens), whereas Blender uses a light ``intensity'' that is determined by a combination of parameters -- including ``power'' (in Watts) which is not equivalent to "power" in the physical sense --  and ``emission strength''. There are other cases in Blender where physical quantities do not have a direct counterpart, so the digital twin must explicitly encode the relationships between physical values and Blender parameters.

    \subsection{Gamma Matching}
        The physical system exhibits a non-linear radiometric (gamma) response, describing the relationship between recorded image intensities and the underlying incident light intensity. Both cameras and projectors typically introduce such non-linearities, which must be compensated to ensure accurate measurements. The gamma correction is commonly expressed as:
        
        \begin{equation}\label{eq:gamma_equation}
            \begin{split}
                I_{out} = a I_{in}^{g} + b
            \end{split}
        \end{equation}
        
        $b$ denotes a constant DC intensity offset, $a$ acts as linear scaling (gain), and $g \ne 1$ denotes the exponent governing a non-linear intensity response (gamma curve).
        
        The obtained gamma parameters are then used to initialise the gamma response of the virtual system in Blender, together with an initial projector radiant intensity $W$.

\begin{figure}[pos=h]
            \centerline{\frame{\includegraphics[width=\columnwidth]{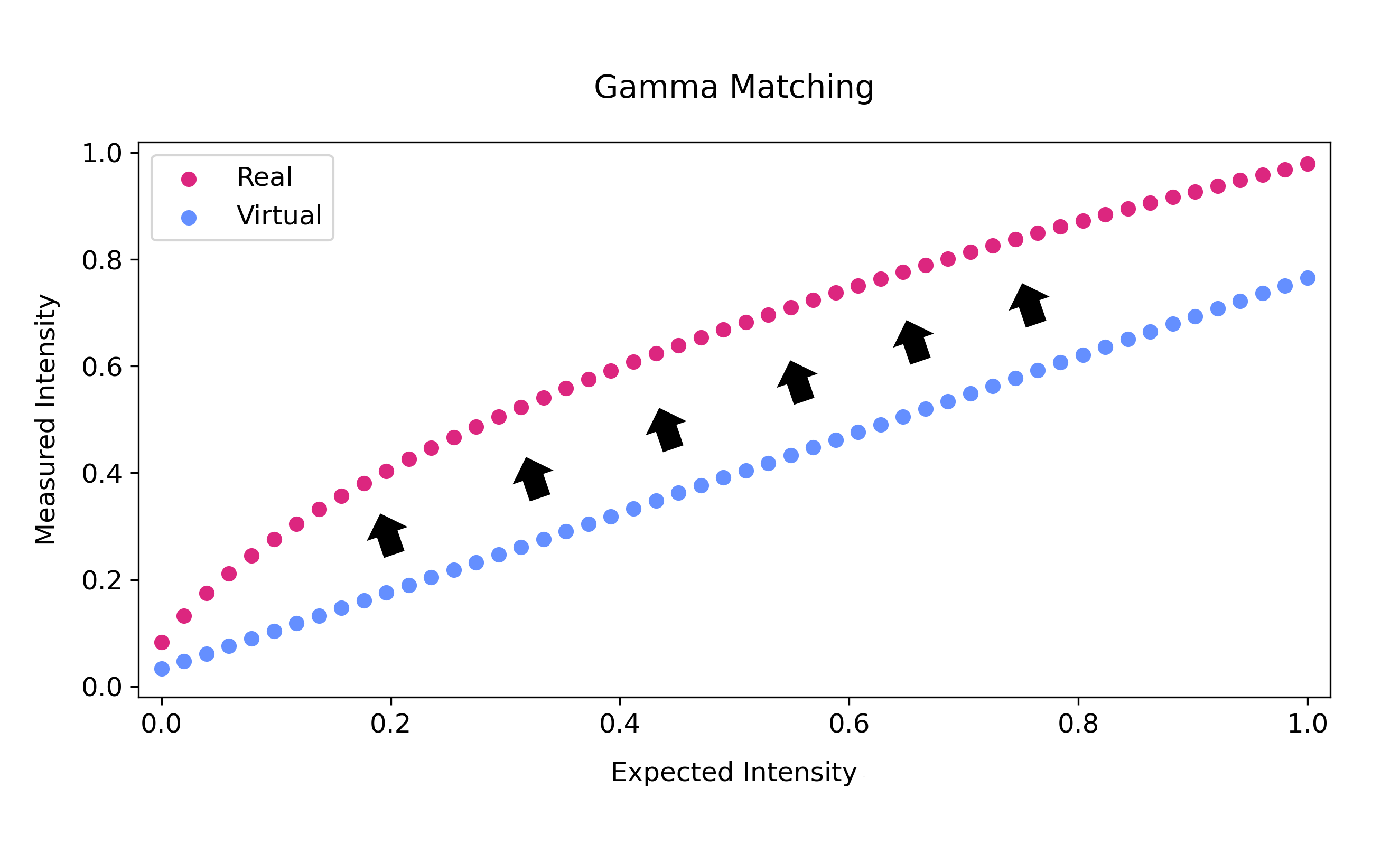}}}
            \caption{The physical system's intensity nonlinearity is matched by the virtual system - in this example, the virtual system becomes more nonlinear as it gets closer to the physical system. $\eqref{eq:gamma_matching}\approx0.002$}
            \label{fig:gamma_match}
        \end{figure}

        The gamma calibration procedure is completed for the virtual system to obtain a set of DC images. The residual mean intensity value for corresponding physical and virtual DC images is calculated to serve as the objective function for a non-linear least-squares optimisation. This optimisation iteratively refines $W_i, a, b, g$ such that the gamma responses of the two systems converge. \eqref{eq:gamma_matching} formalises this non-linear least-squares process, which minimises the discrepancy between the two sets of intensity measurements. One single epoch of this process is represented in \autoref{fig:gamma_match}.

        \begin{equation}
            \label{eq:gamma_matching}
            \begin{split}
                \min_{W, a, b, g} \sum_{i=0}^{N} \left[ x_i - \hat{x}_i(W_i, a, b, g) \right]^2
            \end{split}
        \end{equation}

    \subsection{Characterisation Matching}
        Once the gamma response of the physical and virtual system has been matched, the virtual characterisation process can begin.

        The camera, projector, and calibration board poses obtained from the physical system represent estimates of their true configurations. Although a low reprojection error ($\epsilon_r$) indicates good calibration quality, it does not guarantee that the physical and simulated systems are perfectly aligned. Consequently, an additional refinement step is required to further reconcile the geometric configurations of the physical and virtual environments. To quantify the similarity between physical and simulated characterisation images, a suitable metric is required. In this work, the structural similarity index (SSIM) \cite{hore_image_2010} is employed due to its computational efficiency, differentiability, and widespread use in optimisation frameworks. The standard implementation of SSIM is adopted.
        
        Gradient-based optimisation is used to iteratively adjust the simulation parameters, with SSIM serving as a measure of similarity between physical and virtual images. In principle, increased image similarity corresponds to improved alignment of the characterisation process. However, in practice, both physical and simulated images often contain large low-intensity regions, which can lead to artificially inflated SSIM values. To achieve more robust matching, SSIM is combined with an additional geometric consistency term based on the difference in RMS reprojection error between the physical and virtual systems, defined as $\epsilon_d = \epsilon_r - \epsilon_v$, where $\epsilon_r$ and $\epsilon_v$ denote the reprojection errors of the physical and virtual systems, respectively.

        The inclusion of  $\epsilon_d$ regularises the refinement procedure by constraining the virtual system to remain consistent with the characterisation quality of the physical system. The combination of these two metrics form the characterisation loss function and is expressed in \eqref{eq:characterisation_loss}.

        \begin{equation}\label{eq:characterisation_loss}
            \begin{split}
                L = \alpha \log(1 + \epsilon_d) + \beta (1 - SSIM)
            \end{split}
        \end{equation}

        The values of $\alpha = 0.2$ and $\beta = 0.8$ were selected to prioritise the image-matching capability over $\epsilon_d$, but should be adjusted on a system-by-system basis. One example is for systems which contain a large amount of random sensor noise. $\beta$ can be reduced for this scenario as random noise averages to a small constant error in SSIM (SSIM would have higher variance and is redundant to be weighted so highly).  

        At each iteration of the refinement procedure, the virtual FPP system is rendered in Blender, and the pose (translation and orientation) of the characterisation board is adjusted by the simulation.

        \begin{figure*}[pos=h]
            \centering{\frame{\includegraphics[width=\textwidth]{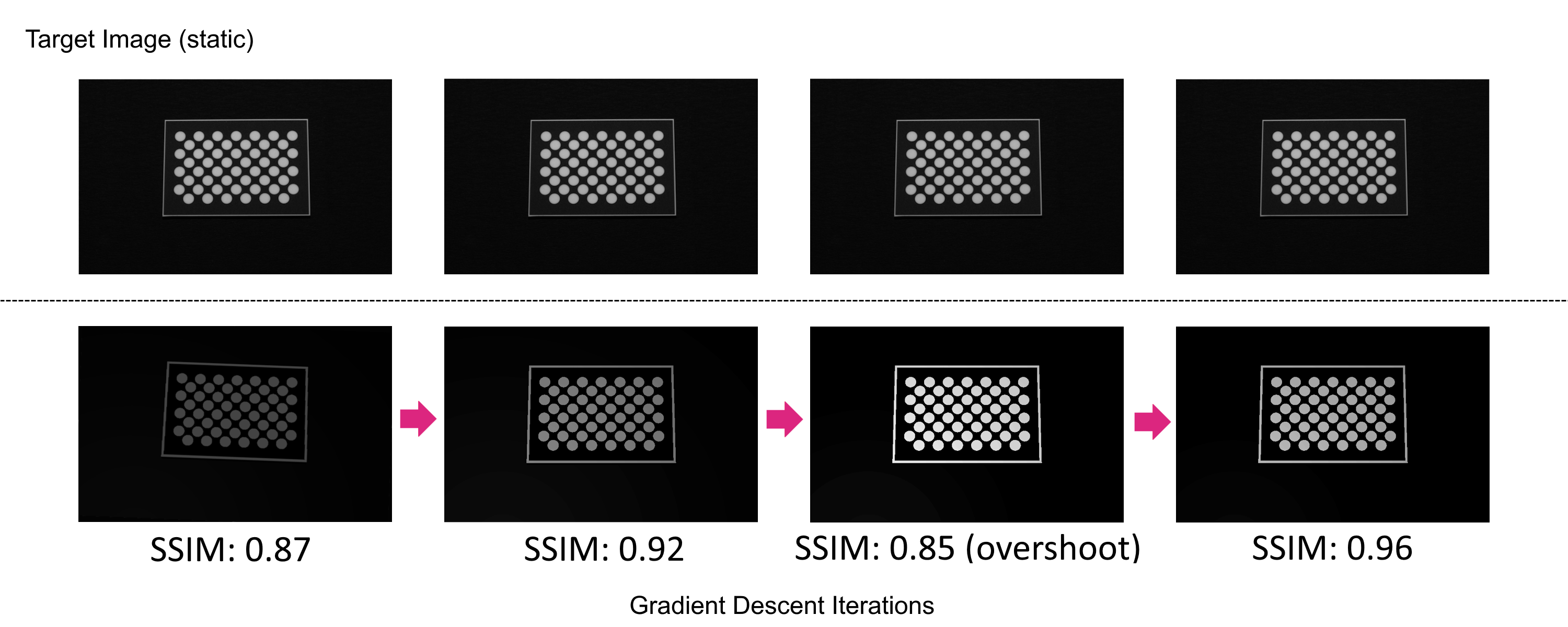}}}
            \caption{Visualisation of digital twin matching: the digital twin optimises the simulation parameters (projector strength, camera gamma response, board transformation), and the loss function converges.}
            \label{fig:converging_blender}
        \end{figure*}

    \subsection{3D Reconstruction Analysis}
\begin{figure}[pos=h]
            \centerline{\frame{\includegraphics[width=\columnwidth]{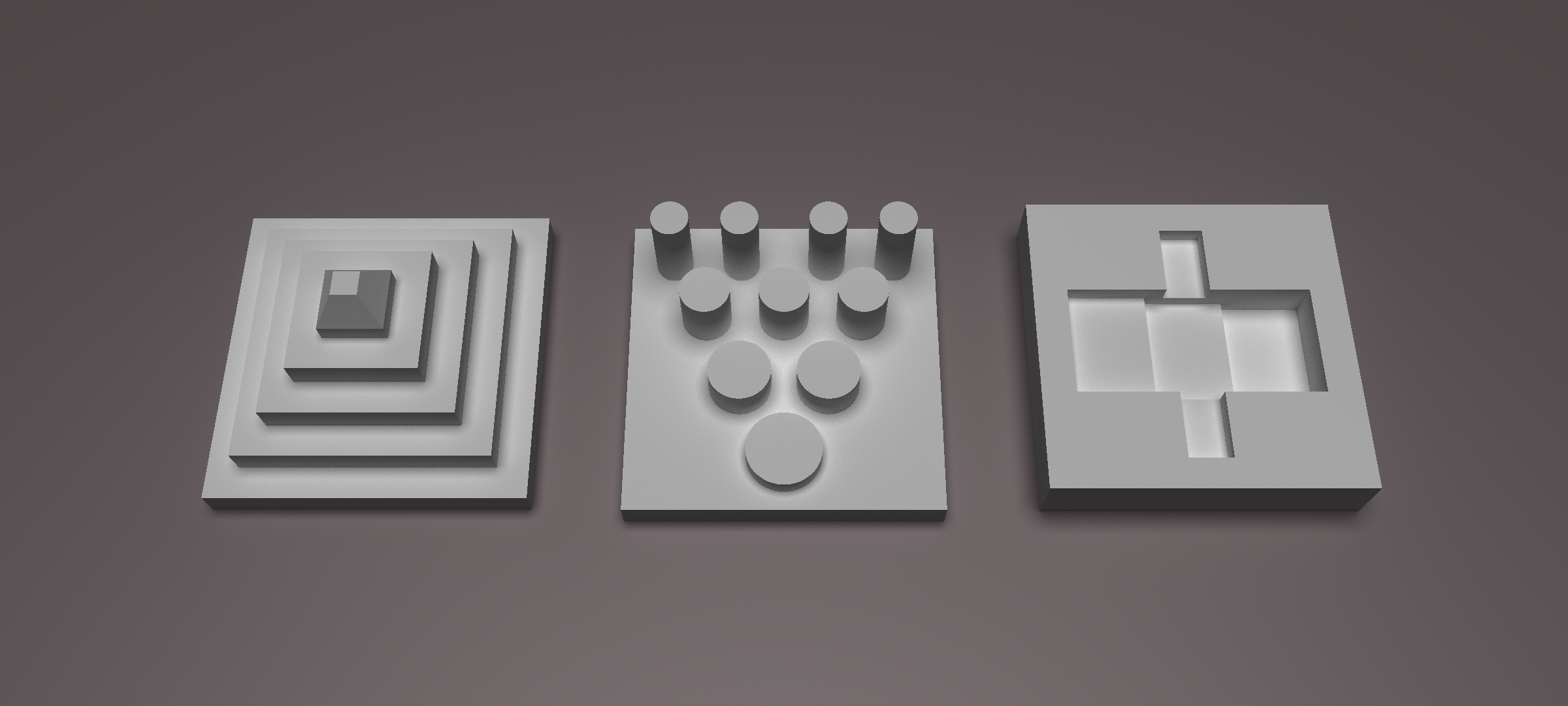}}}
            \caption{Characterisation artefacts: 3D printed pyramid, pillars, and steps}
            \label{fig:virtual-artefacts}
        \end{figure}

        Following characterisation and gamma matching, the system is ready for 3D reconstruction, and the aligned virtual model can be regarded as a digital twin of the physical setup. Reconstruction performance is evaluated using the artefacts shown in \autoref{fig:virtual-artefacts}, which were both simulated in Blender and fabricated via FDM-based 3D printing for physical measurements. As FDM-based 3D printing introduces geometric deviations due to manufacturing parameters (e.g.\ layer height and wall thickness), the sliced digital models -- rather than nominal CAD geometries -- of the artefacts were used in the simulations. This ensures closer consistency between the simulated meshes and the corresponding physical specimens.

        As both the physical and virtual reconstruction processes produce point clouds, a quantitative metric is required to assess their agreement. In this work, the symmetric mean Chamfer distance (SMCD) \cite{akmal_butt_optimum_1998} is used, a standard measure of point-cloud and mesh dissimilarity. SMCD is defined as the mean distance from each point in point cloud $A$ to its nearest neighbour in point cloud $B$ \eqref{eq:mcd}, symmetrised over both sets. Lower SMCD values indicate greater similarity between the point clouds and therefore improved reconstruction accuracy.

        \begin{equation}\label{eq:mcd}
            \begin{split}
                mcd(A, B) = \frac{1}{|A|} \sum_{\mathbf{a} \in A} \min_{\mathbf{b} \in B} \|\mathbf{a} - \mathbf{b}\|_2
            \end{split}
        \end{equation}
        \begin{equation}\label{eq:smcd}
            \begin{split}
                    smcd(A, B) = \frac{mcd(A, B) + mcd(B, A)}{2}
            \end{split}
        \end{equation}

        As we used three measurements artefacts, it is necessary to take the average SMCD value across the three. It is important to note that measurements objects with certain features could be chosen to focus on certain desirable properties to investigate.

        \begin{figure*}[pos=h]
            \centering
            \frame{
                \subfloat{\includegraphics[width=\columnwidth]{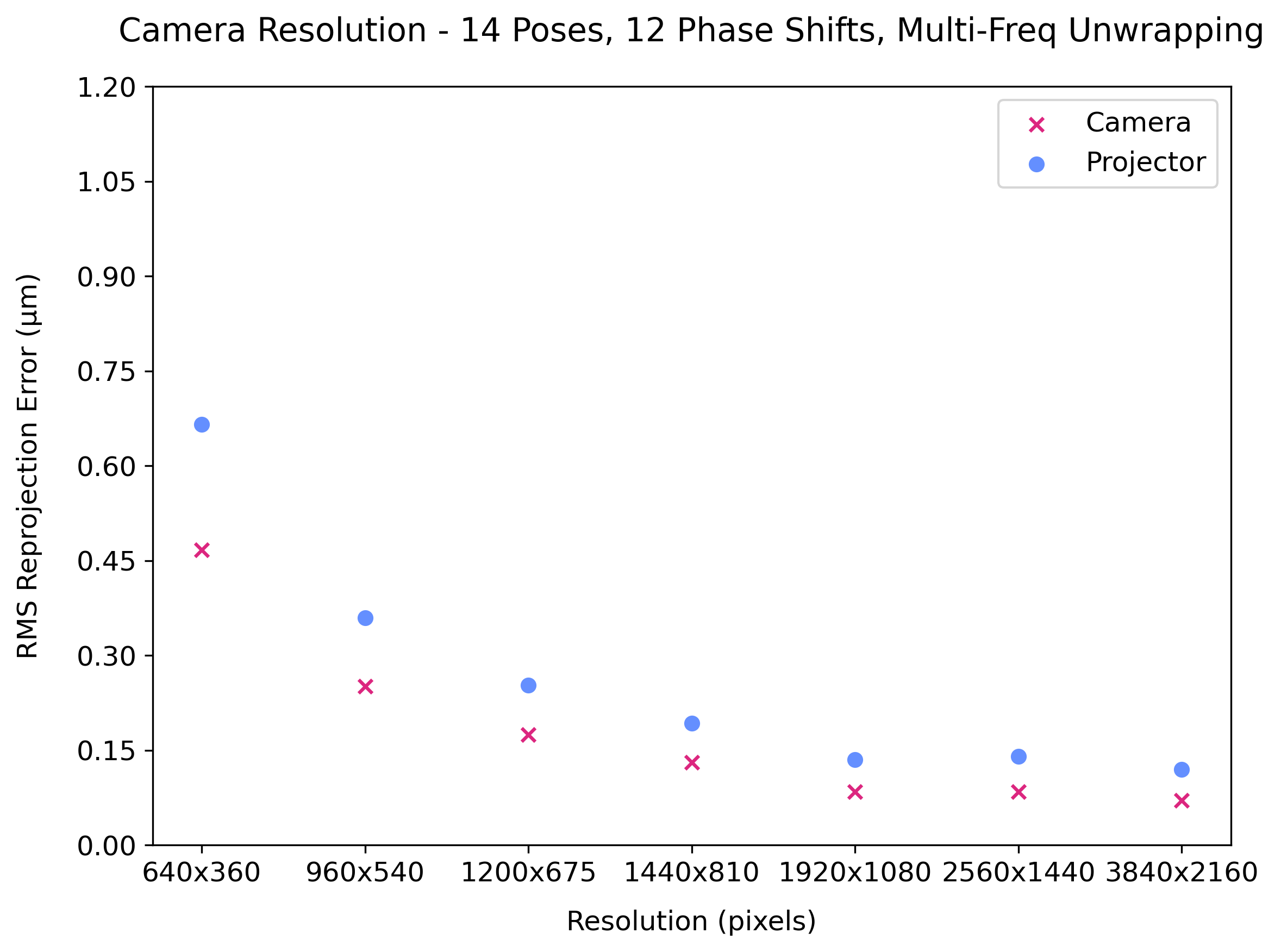}}
                \subfloat{\includegraphics[width=\columnwidth]{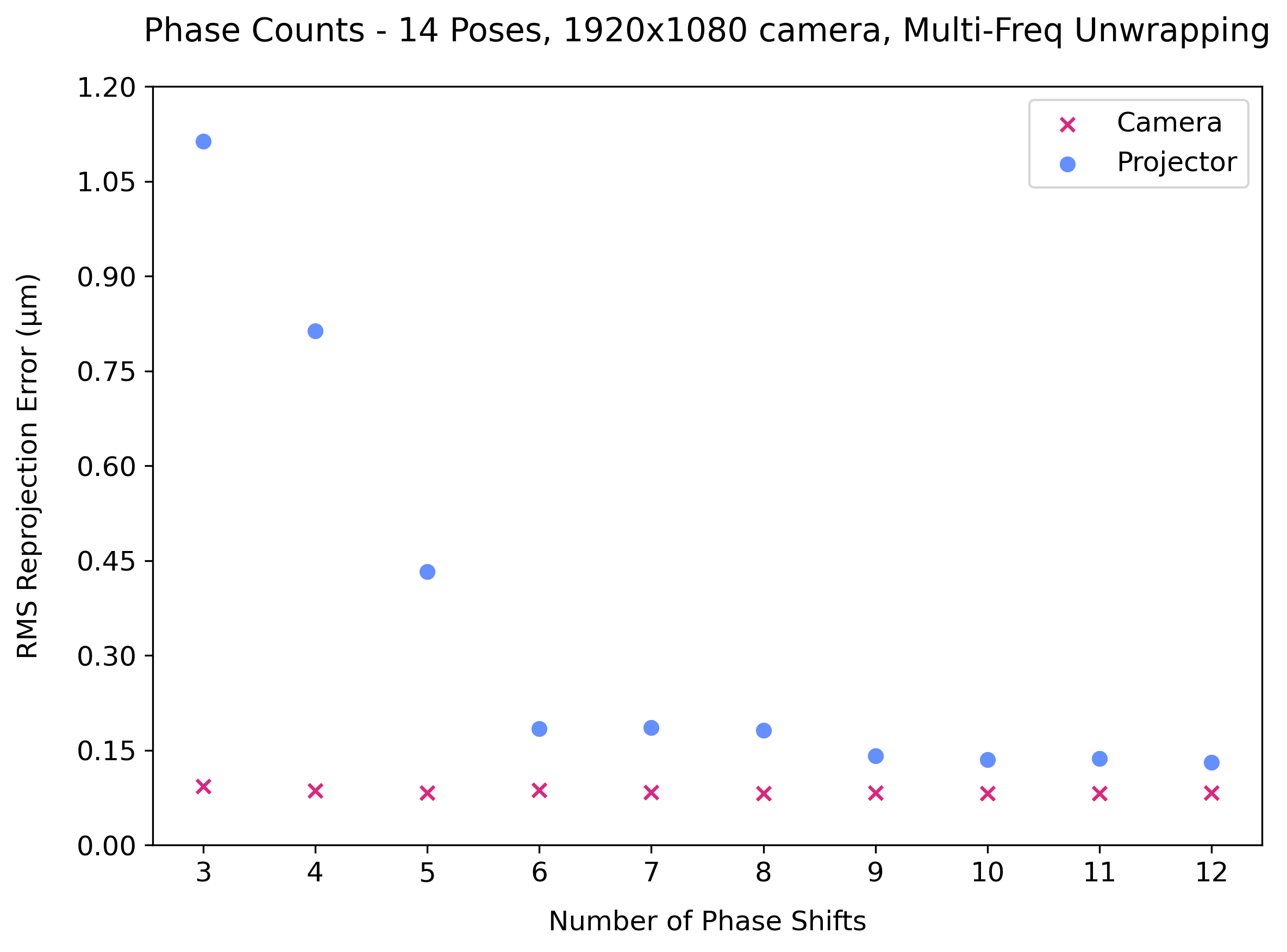}}
            }
            \caption{Sample simulations exploring the effect of increasing camera resolution and number of phases for a purely virtual system. $\epsilon$ is incredibly low compared to physical systems, likely due to low noise presence. The number of stripes used for multi-frequency phase unwrapping was 1, 8, and 64 for the horizontal and vertical fringe directions.}
            \label{fig:overview}
        \end{figure*}

    \subsection{Optimisation}
        With a digital twin established and a mechanism for quality analysis of measurement, we developed an optimisation framework to use the digital twin to enhance the performance of it's mimicked physical system. Optimising parameters in the digital twin domain should; with high likelihood of success; be applicable to the physical system and result in a performance increase. 

        Minimising the mean SMCD of the three artefact measurements to their ground truths (in the digital twin domain) should result in increased reconstruction performance in the physical system. For our work, we made use of a simple gradient descent routine where the mean SMCD metric acts as the loss function, as system improvement was evaluated primarily in terms of reconstruction quality. The termination criterion for the optimiser was triggered if the SMCD did not decrease by \SI{1}{\micro\metre} for five epochs.
        
        In the context of FPP, there are a multitude of parameters which could be explored. For our scenario, we decided to investigate:

        \begin{itemize}
            \setlength\itemsep{0.75em}
            \item Adjustment of the number of phase-shifted images to reduce computational complexity.

            \item Altering the number of stripes per set of fringe patterns used for the temporal multi-frequency scheme.

            \item Adjustment of the baseline spacing between the camera and projector.
        \end{itemize}

        Adjustment of the number of phase-shifted fringe patterns can be performed in conventional FPP systems without the use of a digital twin. However, evaluating whether the digital twin can reproduce this behaviour provides an important proof of principle for the proposed framework. Different phase-shift counts are associated with different fringe frequencies used in temporal phase unwrapping, with higher frequencies typically requiring a greater number of phase shifts to maintain phase estimation accuracy and robustness to noise. To improve computational efficiency, the optimisation can exploit redundancy between phase-shift sets. Specifically, phase-shifting schemes with higher step counts inherently contain subsets corresponding to lower step counts (e.g.\ a 12-step sequence contains all images required for a 6-step sequence). Leveraging this overlap avoids unnecessary re-rendering of images, significantly reducing the computational cost of simulations in Blender.

        The adjustment mechanism for temporal multi-frequency phase unwrapping operates analogously to that used for determining the phase shift count, but involves modification of the number of stripes in the vertical and horizontal directions. Because temporal phase unwrapping inherently requires the projection of a single stripe at the base frequency, this parameter was excluded from the optimisation process. 

        Due to several practical constraints, the baseline distance between the camera and the projector in the physical system could not be investigated in the physical system. Achieving high-precision measurement of the camera and projector positions requires meticulously controlled experimental setup. 

        In this context, the use of a digital twin is particularly advantageous, as it removes the need for any physical intervention, such as disassembling a fixed hardware arrangement. However, this means for our scenario, we were unable to verify whether the parameters produced by the digital twin were valid for the physical system. The camera and projector in the digital twin were positioned in the same positions as specified by their joint characterisations. The digital twin can translate the camera or projector along their common axis in order to increase or decrease their spacing.

\section{Results}
    \subsection{Physical System}
\begin{figure}[pos=h]
            \centerline{\frame{\includegraphics[width=\columnwidth]{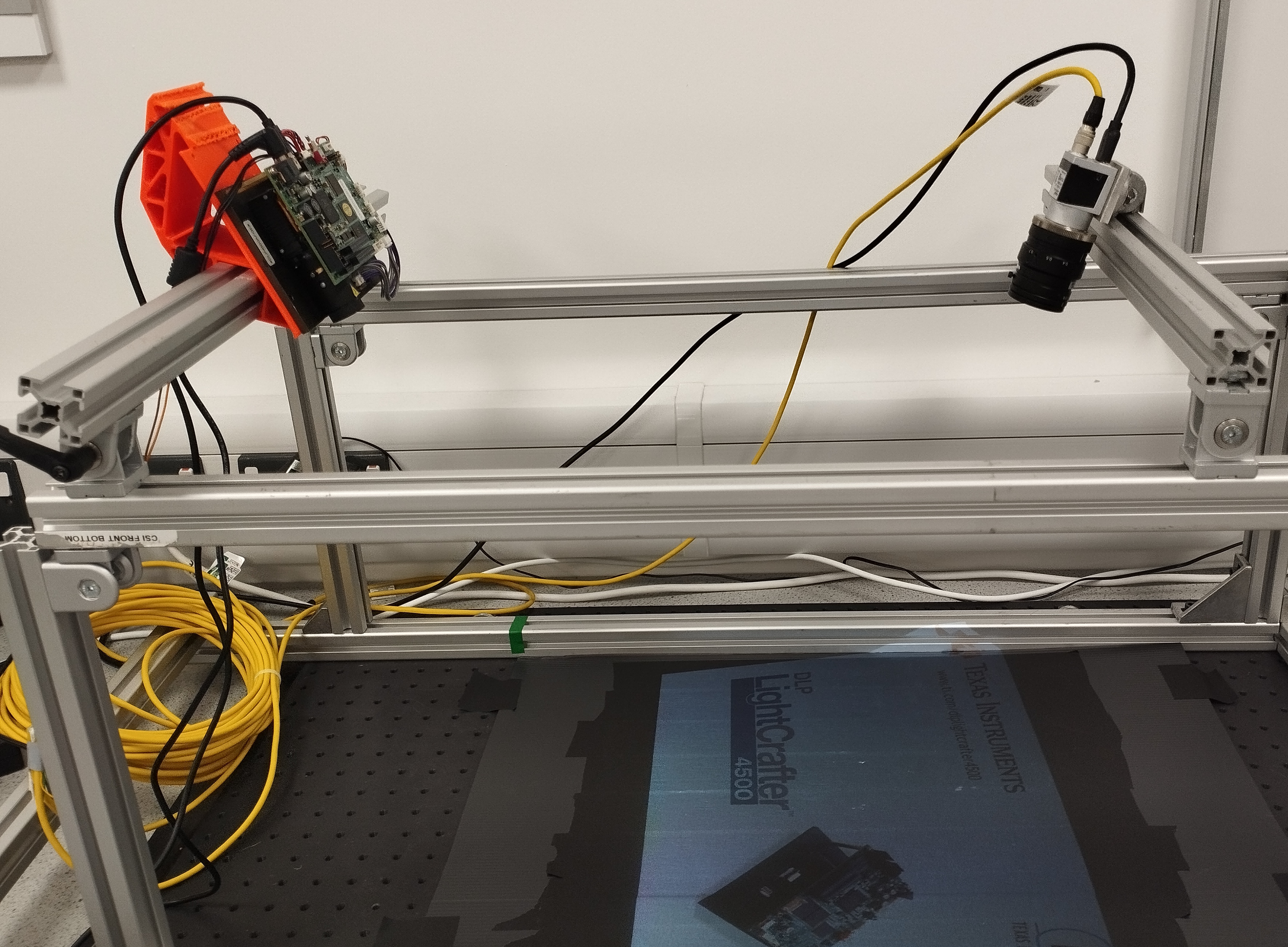}}}
            \caption{The physical metrology system: TI DLP4500 projector, Basler Ace acA5472-17um camera.}
            \label{fig:real_fpp_system}
        \end{figure}

        To test the viability of our digital twin framework a physical FPP setup was used. A DLP4500 lightcrafter projector and Basler Ace camera were deployed. An image of the system can be seen in \autoref{fig:real_fpp_system}. The multi-frequency phase unwrapping method was used with stripe counts of 1, 8 and 64, along with the 12-step phase shifting method.

        We completed characterisation of the physical system using Zhang's method \cite{zhang_camera_2021}. A total of 20 poses of a circle-based characterisation board (see \autoref{fig:real_board}) with varying rotations and translations were used. Reprojection errors of 0.042 (camera) and 0.060 (projector) pixels were achieved which aligns with high quality characterisations.

        The camera and projector individual characterisations were refined using the joint characterisation process \cite{zhang_camera_2021}, obtaining a value of $\epsilon_r = 0.054$. The gathered board poses along with the initial camera and projector poses, formed the complete set of geometric information to approximate the physical system geometrics within the simulation environment.

        Gamma calibration was completed obtaining the values $a=0.65, b=0.056, g=1.033$ -- forming the initial parameters for gamma matching.

    \subsection{Digital Twin Matching}
\begin{figure}[pos=h]
            \centerline{\frame{\includegraphics[width=\columnwidth]{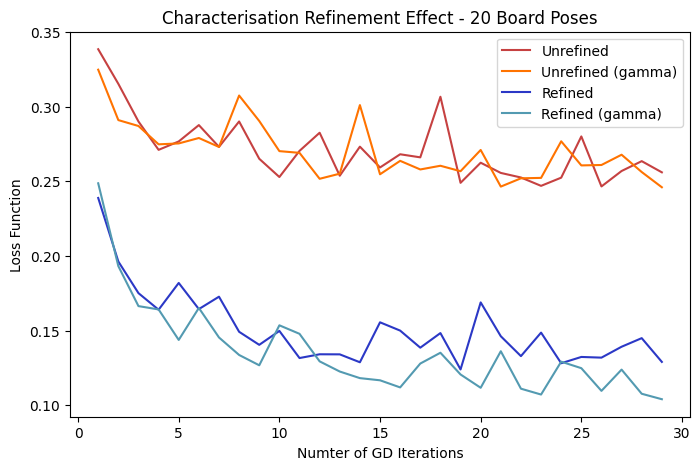}}}
            \caption{$L \rightarrow ~0.002$ after around 30 epochs. The standard deviation of $L$ tends to vary across images - a point on the characterisation board may become slightly quantised at a particular pose, or less visible due to increased environmental lighting influence.}
            \label{fig:char_loss}
        \end{figure}

        We completed the characterisation matching process for our physical system, and \autoref{fig:char_loss} shows the loss function converging to zero after approximately 25 epochs.

        As explained earlier, reprojection error is only a heuristical estimate of the performance of charactersation as it is prone to overfitting. Our loss function tries to combat this by also matching images directly, and the converging system performance in \autoref{fig:char_loss} can be supported by manual verification (such as \autoref{fig:converging_blender}).

        \begin{table}[pos=h]
            \centering
            \begin{tabular}{|l|l|l|l|}
            \hline
            \multicolumn{1}{|c|}{Characterisation Refined?} & \multicolumn{1}{c|}{SSIM} & \multicolumn{1}{c|}{$\epsilon_d$} & \multicolumn{1}{c|}{$L$} \\ \hline
            No                               & 0.701 & 0.081 & 0.256 \\ \hline
            No (with gamma correction):      & 0.713 & 0.072 & 0.246 \\ \hline
            Yes                              & 0.841 & 0.007 & 0.129 \\ \hline
            Yes (with gamma correction):     & 0.872 & 0.002 & 0.104 \\ \hline
            \end{tabular}
            \caption{\centering SSIM and $\epsilon_d$ values produced by our physical system after refinement in Blender. The gamma calibration process is essential for a higher SSIM score.}
        \end{table}

\begin{figure}[pos=h]
            \centerline{\frame{\includegraphics[width=\columnwidth]{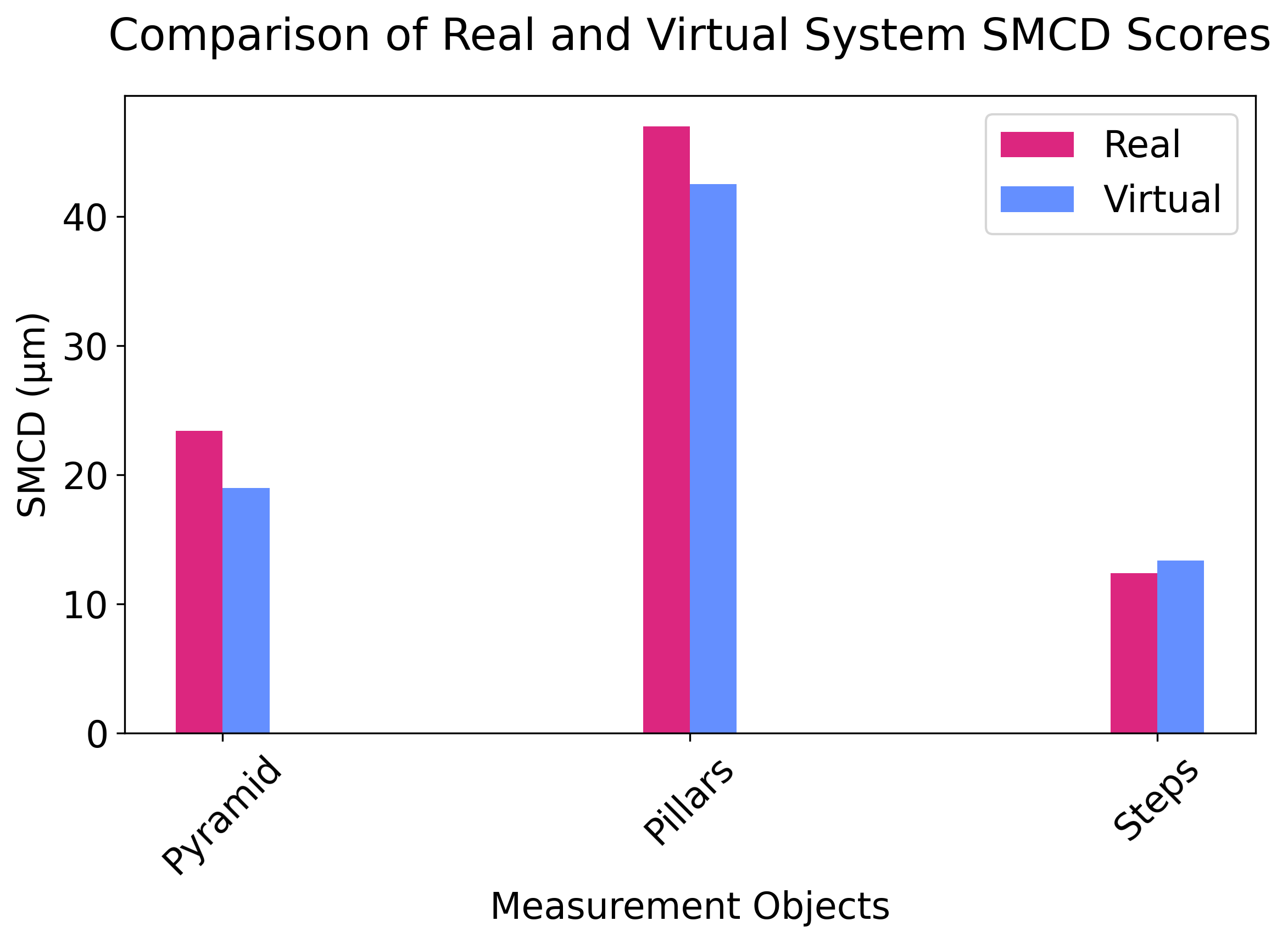}}}
            \caption{The steps artefact performs similarly across the physical and virtual system, whereas the pillars and pyramid artefacts display disparity. This could be explained by local regions of shadowing being less prevalent in the steps artefact than the others.}
            \label{fig:smcd_comparison}
        \end{figure}

        The SMCD was compared across the two systems for three artefacts (see \autoref{fig:virtual-artefacts}) and is presented in \autoref{fig:smcd_comparison}. The virtual system performs better for measuring artefacts in each of the three scenarios. One possible explanation is that the Blender simulation produces perfectly in-focus images with infinite depth of field, allowing for slightly better performance across the whole measurement volume. This results in less variance across local regions of the object with respect to SMCD. It could also be due to the virtual environment not being sufficiently modelled to match the physical environment (with respect to lighting and geometrics).

        The average $\Delta$SMCD was \SI{13}{\micro\metre} for the three artefacts in \autoref{fig:smcd_comparison}; the virtual system performed around 19\% better on average.

    \subsection{Optimisation}
        \begin{table*}[pos=t]
            \centering
            \small
            \setlength{\tabcolsep}{4pt}
            \renewcommand{\arraystretch}{1.25}
            \begin{adjustbox}{max width=\textwidth}
                \begin{tabular}{|c|c|c|c|c|c|c|}
                    \hline
                    \textbf{Parameter} & \textbf{System} & \textbf{Initial Value} & \textbf{Optimised Value} & \textbf{Epochs} & \textbf{Initial Mean SMCD} & \textbf{Optimised Mean SMCD} \\
                    \hline

                    \multirow{2}{*}{\makecell{Phase Counts}} 
                    & Physical 
                    & \multirow{2}{*}{\makecell{(12, 12, 12)}}
                    & (6, 6, 9)
                    & 21
                    & \SI{31.2}{\micro\metre} 
                    & \SI{31.0}{\micro\metre} \\
                    \cline{2-2}\cline{4-7}
                    & Digital Twin 
                    & 
                    & (6, 6, 9)
                    & 17 
                    & \SI{28.1}{\micro\metre} 
                    & \SI{24.9}{\micro\metre} \\
                    \hline

                    \multirow{2}{*}{\makecell{Stripe Counts}} 
                    & Physical 
                    & \multirow{2}{*}{\makecell{(1.0, 4.0, 16.0)}}
                    & \makecell{(1.0, 8.20, 113.75)}
                    & 55 
                    & \SI{94.1}{\micro\metre} 
                    & \SI{24.5}{\micro\metre} \\
                    \cline{2-2}\cline{4-7}
                    & Digital Twin 
                    & 
                    & \makecell{(1.0, 9.43, 142.05)}
                    & 64 
                    & \SI{73.3}{\micro\metre} 
                    & \SI{24.3}{\micro\metre} \\
                    \hline

                    \makecell{Camera--Projector\\Spacing}
                    & Digital Twin 
                    & \SI{1.20}{m} 
                    & \SI{1.132}{m} 
                    & 12 
                    & \SI{31.2}{\micro\metre} 
                    & \SI{19.7}{\micro\metre} \\
                    \hline
                \end{tabular}
            \end{adjustbox}
            \caption{The initial camera--projector spacing was determined to be suboptimal by the optimisation routine. A smaller spacing was introduced, suggesting that shadowing was dominating the measurement quality.}
            \label{tab:optimisation_results}
        \end{table*}

        Our digital twin was able to find a minima for the number of phase shifts with respect to computational runtime. It was able to reduce the total number of projected fringes down from 36 to 21, resulting in a ~48\% decrease in the number of images per reconstruction. This result could be particularly useful for dynamic fringe projection as these systems tend to require very low-latency measurements (utilising a global shutter camera) in order to reduce motion blur -- Blender does support the animation of objects with motion blur.

        The mechanism for adjusting the temporal multi-frequency phase unwrapping works in the same manner as for phase shift count, but by adjusting different parameters. Projecting a single stripe for the base frequency is a requirement for temporal phase unwrapping, and was not included in the optimisation parameters. The y-direction fringes were only used for reconstruction; x-direction stripes are redundant as the x-coordinate of points in the reconstructed cloud can be calculated using only the y-direction phasemap.

        The purpose of running the optimisation for both the physical and virtual systems is to compare how their measurement performance is after optimisation. Large discrepancies indicate that the systems are misaligned and hence a poor digital twin is developed.
        The physical system reached equilibrium after a longer number of epochs compared to the virtual system, and resulted in a slightly better SMCD score. 

        This indicates that the virtual system had more ``potential'' for optimisation than the physical system, possibly due to less noise present in the measurements. We investigated the cause of this, and it was determined that the physical system contained more measurement outliers than the virtual system (a larger variance in per-point SMCD score). This increased variances inflates the SMCD score, but is amongst a common encountering in 3D measurement analysis. Filtering; and a robust shadow filter method\footnote{compared to our AC-DC masking}; could be deployed to mitigate the effect of outliers influencing SMCD.

        As a result of optimisation, the physical system moved the camera and projector closer together to achieve a better SMCD score. Theoretically, this decreases the triangulation angle and should decrease the quality of depth resolution. However, it is important to recognise with large baseline-distances, shadowing is more likely to occur and create large outliers is they are not sufficiently filtered. Trivially, the optimisation did not move the devices extremely close as the triangulation angle would be affected. Logically, there exists some baseline distance for which the optimal position lies; \cite{wu_optimization_2024} explores this theoretically and loosely aligns with our findings.

\section{Discussion}

We have demonstrated a highly effective Blender-based framework for creating digital twins of fringe projection profilometry (FPP) metrology systems. The proposed system can closely match physical system performance and support closed-loop optimisation, as validated by comparing reconstructed mesh accuracy against ground truth using the mean Chamfer distance. In addition to supporting the optimisation and refinement of existing systems through digital twin methodologies, it also enables virtual prototyping of FPP setups without physical hardware.

Several limitations of the current implementation warrant further investigation. First, the optimisation strategy employed in this study relies on straightforward gradient descent to guide system improvements. This restricts the pipeline to single-objective optimisation and limits its ability to balance competing performance metrics. Future work could incorporate multi-objective optimisation methods, such as Pareto-front approaches, to enable more nuanced trade-offs between parameters. For example, optimising the projector--camera baseline and the phase-shift count independently may lead to recommendations that are suboptimal when combined: reducing the baseline may improve resolution, while simultaneously reducing the number of phase shifts may increase sensitivity to noise. Without jointly considering such interactions, the digital twin may produce conflicting or fragile configurations. Multi-objective optimisation would allow these trade-offs to be explored explicitly and would support more robust system design.

A second limitation concerns the generalisability of the digital twin to more complex real-world environments. The system developed in this study was validated against a relatively simple physical scene: a flat object with minimal environmental clutter, representative of a typical manufacturing metrology setting. In more geometrically complex or cluttered environments, such as some manufacturing facilities, replicating the physical scene in Blender may become significantly more challenging. Accurate modelling of background geometry, material properties, and lighting interactions requires expertise in both FPP systems and Blender's rendering pipeline. Suboptimal environmental modelling could lead to slower convergence during system refinement, or even failure to achieve a valid calibration. Nevertheless, for professional-grade metrology systems, the assumption of a relatively simple physical scene and background is often reasonable.

Despite these limitations, the digital twin framework developed here has strong potential to support the integration of artificial intelligence in FPP, and in optical metrology more broadly. AI-based methods, particularly deep learning, are increasingly being adopted in structured-light systems. For instance, convolutional neural networks have been shown to reconstruct 3D shape from a single fringe pattern, enabling real-time measurement of dynamic scenes \cite{wang_deep_2023}. Recent advances have demonstrated frame rates on the order of 100{,}000 fps using deep neural networks to map low-quality input fringes to high-quality phase maps \cite{wang_single-shot_2025}. These approaches offer major advantages for dynamic or high-throughput applications, but they also face well-documented challenges, including limited generalisation and the need for large annotated training datasets. In FPP, acquiring large quantities of ground-truth 3D data from real-world scenes is often laborious or infeasible, which limits the practicality of supervised learning. To address this, several studies have proposed generating synthetic datasets using virtual FPP systems built in graphics engines such as Blender. These simulations can produce large volumes of paired fringe and phase data under controlled conditions, substantially reducing the cost of data acquisition while improving training generalisability \cite{zhou_synthetic_2022}. The digital twin system presented in this paper could therefore serve as a foundation for generating such datasets, helping to accelerate the development of AI-driven FPP algorithms.

\section{Acknowledgements}
    GSDG would like to acknowledge support from UKRI (MR/Y034163/1 and MR/Z505183/1).
    DW acknowledges fundings from an EPSRC Studentship (2879031).
    SP acknowledges support from an EPSRC Grant (EP/X010929/1).

\section{Bibliography}

\bibliographystyle{bibstyle1} 

\bibliography{references}

\end{document}